\documentclass[journal,comsoc]{IEEEtran}
\usepackage[T1]{fontenc}% optional T1 font THIS encoding
\usepackage{times}  % DO NOT CHANGE THIS
\usepackage{helvet} % DO NOT CHANGE THIS
\usepackage{courier}  % DO NOT CHANGE THIS
\usepackage{euscript}
\usepackage[hyphens]{url}  % DO NOT CHANGE THIS
\usepackage{graphicx} % DO NOT CHANGE THIS
\usepackage{amsmath,amssymb}
\usepackage{mathtools}

\usepackage[ruled, vlined]{algorithm2e}
\usepackage{multirow}
\usepackage{booktabs}

\begin{document}

\title{Topology and Content Co-Alignment Graph Convolutional Learning}

\author{Min~Shi,~\IEEEmembership{Student Member,~IEEE,}
        Yufei~Tang,~\IEEEmembership{Member,~IEEE,}
        and~Xingquan~Zhu,~\IEEEmembership{Senior~Member,~IEEE}% <-this % stops a space
        \thanks{This work was supported in part by the US National Science Foundation (NSF) through Grants No. IIS-1763452 \& CNS-1828181.}
\thanks{M. Shi, Y. Tang, and X. Zhu are with the Department of Computer \& Electrical Engineering and Computer Science, Florida Atlantic University, Boca Raton, FL 33431 USA. E-mail: \{mshi2018, tangy, xzhu3\}@fau.edu.} }

% The paper headers
% \markboth{IEEE Transactions on Neural Networks and Learning Systems, March~2020}%
% {Shell \MakeLowercase{\textit{et al.}}: Bare Demo of IEEEtran.cls for IEEE Communications Society Journals}

\maketitle

\newcommand{\nickname}{CoGL}

\newcommand{\graph}{\text{G}}
\newcommand{\nodes}{\textbf{V}}
\newcommand{\edges}{\textbf{E}}
\newcommand{\feats}{\textbf{X}}
\newcommand{\nodenum}{|\textbf{V}|}
\newcommand{\edgenum}{|\textbf{E}|}
\newcommand{\adjmat}{\textbf{A}}
\newcommand{\adjmattilde}{\tilde{\adjmat}}
\newcommand{\degmat}{\textbf{D}}
\newcommand{\idenmat}{\textbf{I}}
\newcommand{\R}{\mathbb{R}}
\newcommand{\embeds}{\text{O}}
\newcommand{\embedbar}{\bar{\textbf{X}}^{(2)}}
\newcommand{\Aij}{\textbf{A}_{i,j}}
\newcommand{\feati}{\textbf{X}_i}
\newcommand{\featj}{\textbf{X}_j}
\newcommand{\nodei}{v_i}
\newcommand{\edgeij}{e_{i,j}}
\newcommand{\wij}{w_{i,j}}
\newcommand{\featdim}{m}
\newcommand{\embedsize}{c}
\newcommand{\adjmatbar}{\bar{\adjmat}}
\newcommand{\lossfeat}{\mathcal{L}_{cont}}
\newcommand{\lossgcn}{\mathcal{L}_{gcn}}
\newcommand{\lossgan}{\mathcal{L}_{gan}}
\newcommand{\Abarij}{\bar{\adjmat}_{i,j}}
\newcommand{\Wc}{\textbf{W}_c}
\newcommand{\Wp}{\textbf{W}_p}
\newcommand{\Wone}{\textbf{W}^{(1)}}
\newcommand{\Wtwo}{\textbf{W}^{(2)}}
\newcommand{\wpdim}{d}
\newcommand{\relu}{\text{ReLU}}
\newcommand{\hiddendim}{h}
\newcommand{\labnodes}{\mathcal{Y}_L}
\newcommand{\embedsoftmax}{\text{H}}
\newcommand{\labonehot}{\text{Y}}
\newcommand{\gen}{\mathcal{G}}
\newcommand{\dis}{\mathcal{D}}
\newcommand{\generator}{\gen(\adjmat,\feats)}
\newcommand{\sample}{x}
\newcommand{\discriminator}{\dis(\sample)}
\newcommand{\E}{\mathbb{E}}
\newcommand{\realdatadist}{\sample\sim \embedbar}
\newcommand{\fakesample}{z}
\newcommand{\fakedatadist}{\fakesample}
\newcommand{\dgnator}{\dis(\generator)}

\begin{abstract}
In traditional Graph Neural Networks (GNN), graph convolutional learning is carried out through topology-driven recursive node content aggregation for network representation learning. In reality, network topology and node content are not always consistent because of irrelevant or missing links between nodes. A pure topology-driven feature aggregation approach between unaligned neighborhoods deteriorates learning for nodes with poor structure-content consistency, and incorrect messages could propagate over the whole network as a result. In this paper, we advocate co-alignment graph convolutional learning (CoGL), by aligning the topology and content networks to maximize consistency. Our theme is to force the topology network to respect underlying content network while simultaneously optimizing the content network to respect the topology for optimized representation learning. Given a network, CoGL first reconstructs a content network from node features then co-aligns the content network and the original network though a unified optimization goal with (1) minimized content loss, (2) minimized classification loss, and (3) minimized adversarial loss. Experiments on six benchmarks demonstrate that CoGL significantly outperforms existing state-of-the-art GNN models.
%The great success of existing Graph Neural Networks (GNN) largely relies on the recursive feature aggregation mechanism where graph features (e.g., text attributes) and structures enhance each other to achieve optimized node representations. However, features and structures from the graph data are not always consistent for potentially incomplete and irrelevant links between nodes. Feature passing between unaligned neighborhoods will cause false representations for some nodes and the incorrect messages could further propagate over the whole structure. In this paper, we propose a novel GNN model, \text{\nickname}, to explore features and structures balancing for optimal and robust graph embedding. \text{\nickname} seeks to reconstruct graph structures (e.g., feature network) from node features, which is then used for adversarial co-training with the original structure network. Another advantage of \text{\nickname} is that embedding learning parameters are shared between the two distinct feature and structure networks to collaboratively derive optimal graph structures and node representations. Experimental results on six benchmarks demonstrate that our approach significantly outperforms existing state-of-the-art GNN models.

\end{abstract}


\section{Introduction}
%Graph Neural Networks (GNN) have recently received significant attention due to the remarkable achievements in many knowledge mining and analytic tasks such as social network mining \cite{fan2019graph} and image recognition \cite{chen2019multi}, \textit{etc}. It is widely acceptable that the great success of GNN attributes to the efficient message passing mechanism which can easily encode node features and graph structures in a unified latent space. The main idea behind \cite{kipf2016semi,hamilton2017inductive} is that nodes can iteratively aggregate features from their respective local graph neighborhoods. Meanwhile, features of each node can be propagated through graph structures to impact other nodes. Hence, both node features and topological structures will be naturally preserved throughout the interactive learning process between features and structures. With such a learning paradigm, in the past few years tremendous efforts have been focused on developing more effective feature aggregators for improved node representation learning, such as GAT \cite{velivckovic2017graph} that learns different importance weights for various neighborhoods and GraphSAGE \cite{hamilton2017inductive} that learns a set of aggregator functions for each node.

Recent years have witnessed a significant growth of Graph Neural Networks (GNN) in solving domain specific tasks such as social network mining \cite{fan2019graph} and image recognition \cite{chen2019multi}, \textit{etc}. This is mainly attributed to GNN's efficient message passing mechanism to encode network topology and node content/features in a unified latent space, through iterative feature aggregation of neighborhoods for each node~\cite{kipf2016semi,hamilton2017inductive}. Because node content (or node features) are propagated through network topology, both node features and topological structures are naturally preserved throughout the interactive learning process between features and structures. Under this learning paradigm, tremendous effort has been focused on developing effective feature aggregators for improved node representation learning, such as GAT~\cite{velivckovic2017graph} that learns different importance weights for various neighborhoods and GraphSAGE~\cite{hamilton2017inductive} that learns a set of aggregator functions for each node.

%Despite of the promising results, existing methods naturally consider that node features and node relationships are aligned and mutually enhanced during the learning \cite{wu2019comprehensive}. However, in real-world scenarios node features (e.g., text attributes and image pixels) and graph structures (e.g., link relations between nodes) from networked data are not always consistent due to some unconscious or deliberate human behaviors \cite{hofmann2017dbkwik,liu2019projfe}. For example, irrelevant citations between different scholarly publications would cause inconsistent citation networks. Similarly, an attacker could create fake followers or manipulate friendship relations on social media, resulting in inconsistent social networks \cite{bojcheski2018adversarial}. Apart from the inconsistent or incorrect structural relations between graph nodes, incomplete or missing links between nodes in a graph are also the common situations, \textit{i.e.}, one might builds an image graph based on the tag sharing rule while the tag information could be sparse or absent for some image nodes. Figure 1 shows two examples to illustrate the cases of inconsistent and incomplete graphs or networks. In summary, the graph inconsistency combined with the sparse node relations would seriously challenge existing GNN learning models for the following two reasons:

\begin{figure}
\begin{small}
  \centering
    \includegraphics[width=0.48\textwidth]{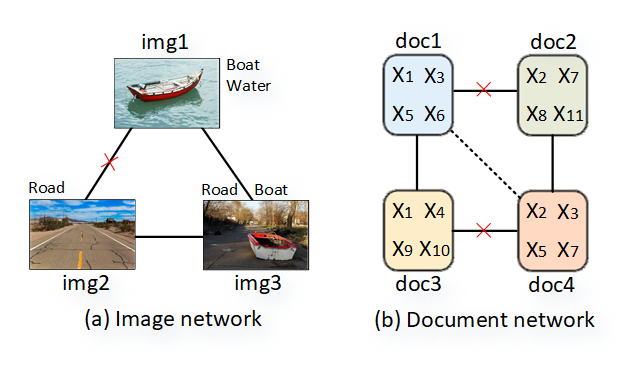}
    \vspace{-6mm}
\caption{An illustration of inconsistent/incomplete relationships in networks. (a) an image network. (b) a document network. Topology of both networks are given but are not always consistent to the node content. For example, when using image semantics as node content, the edge between img1 and img2 is inconsistent as they do not share any common semantics (they might be connected because of other unobserved correlations, such as pictures taken during the same trip). Analogically, the inconsistencies exist between (doc1, doc2) and (doc3, doc4), and there is a potential missing edge between doc1 and doc4 due to common features X$_3$ and X$_5$.}%  Label correlations should be modeled in multi-label graph learning.}  
  \label{fig:motivation}
  \end{small}
%  \vspace{-4mm}
\end{figure}

Despite of promising results, all existing graph convolutional learning approaches employ a topology-driven principle to ``force'' node content to be aggregated by following edge connections. By doing so, they consider that node features and edge relationships are largely consistent and are mutually enhanced during the learning \cite{wu2019comprehensive}. In reality, network node content (\textit{e.g.}, text and image semantics) and graph topology (\textit{e.g.}, link relations) may be highly inconsistent due to unconscious or deliberate human behaviors~\cite{hofmann2017dbkwik,liu2019projfe}. For example, irrelevant citations between scholarly publications result in inconsistent citation networks. Similarly, an attacker may create fake followers or manipulate friendships, resulting in inconsistent social networks~\cite{bojcheski2018adversarial}. In addition, incomplete or missing links between nodes are also common, \textit{i.e.}, an image graph based on tag sharing rules often have sparse tag information therefore results in missing edges. Fig.~\ref{fig:motivation} shows two examples of inconsistent and incomplete graphs or networks. In summary, graph inconsistency combined with sparse node relations severely challenge existing GNN learning models for following two reasons:
\begin{itemize}
    \item \textbf{Noisy Message Passing:} When neighborhood relationships are misaligned to node affinities reflected by node content, nodes will aggregate irrelevant information from neighbors, resulting in noisy content and inferior representation learning. Such noisy messages will continue to pass through graph structures and finally deteriorate the learning of all nodes.
    \item \textbf{Node Relationship Impairing:} When two nodes have similar node content but there is no link between them, the missing link will ``force'' similar nodes to be dissimilar in embedding space. In addition, these absent relations will further impair the relation modeling between other pairs of nodes over the entire network.
%    \item \textbf{Lack Relation Constrain:} If features act to reveal high similarity between two nodes but there is no corresponding direct link observed, the structural constrain could thus be weaken between them. In addition, these absent relations would further impair the relation modeling between other pairs of nodes over the whole graph.
\end{itemize}

\begin{figure}
\begin{small}
  \centering
    \includegraphics[width=0.45\textwidth]{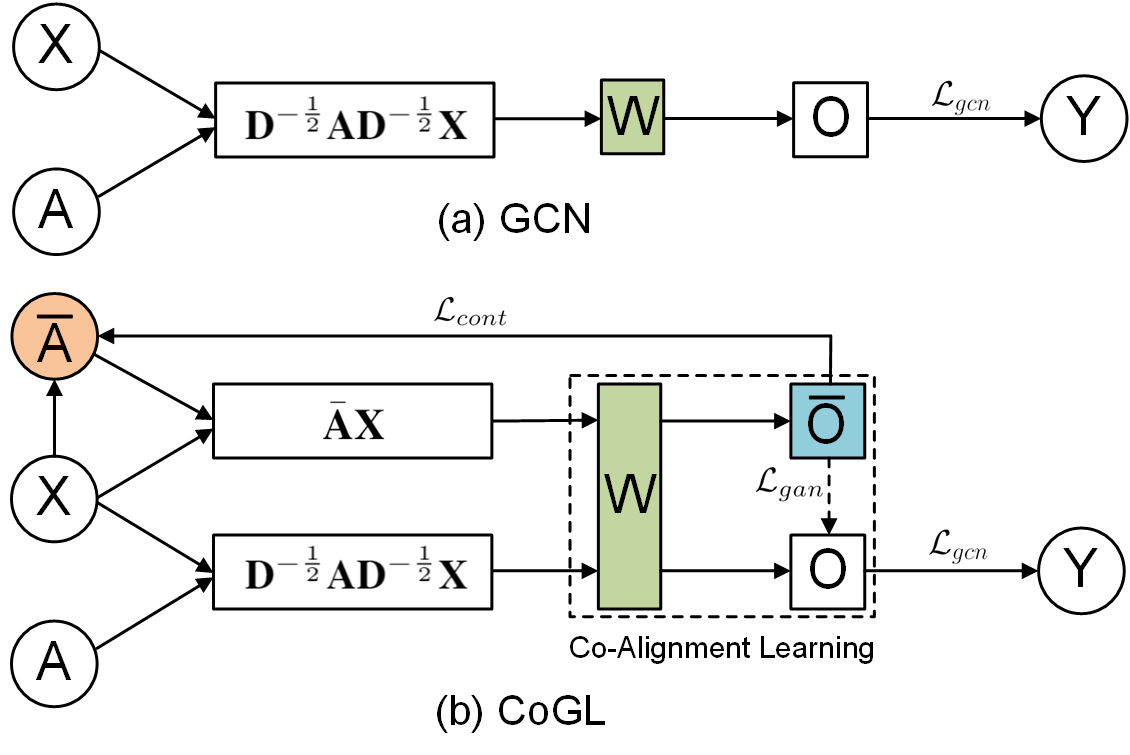}
    \vspace{-4mm}
\caption{The difference between GCN and our proposed CoGL. (a) GCN learns graph embeddings based on network topology $\adjmat$ and node content $\feats$ by optimizing classification loss $\lossgcn$. (b) CoGL first learns to construct a content network $\adjmatbar$ and then performs content and topology co-alignment embedding learning by optimizing a unified loss from three parts: content network construction loss $\lossfeat$, node classification loss $\lossgcn$ and adversarial training loss $\lossgan$.}%.}  
  \label{fig:diff}
  \end{small}
%  \vspace{-4mm}
\end{figure}

Notice that network topology is often not optimal, a recent work~\cite{Yang2019topology} proposes to optimize network topology structure to improve graph convolutional network learning. However, such approach is potentially risky because revising network topology to satisfy optimization tends to overfit to training data. Alternatively, in this paper, we propose taking a different route to co-align network topology and node content for graph convolutional learning.

%the discrepancy between graph features and structures into considertion for optimal and robust networked data learning. 
Specifically, we focus on information networks where nodes have rich features (content) such as texts and images. Instead of employing the traditional topology-driven principle, we treat network topology and node content as two distinct, but highly correlated, sources of data, and explicitly characterize their differences for embedding learning. We advocate a new ``co-alignment'' learning principle where the topology network should respect underlying node content, and the content should comply to the topology network for optimized representation learning. We propose a GNN model namely \textit{\textbf{Co}-alignment \textbf{G}raph convolutional \textbf{L}earning} (\text{\nickname}) for this purpose. 

To enable co-alignment learning, \text{\nickname} reconstructs a content network from node features. The content network together with the original topology network are then set to perform co-alignment learning in an adversarial fashion: (1) the content network aims to learn good embeddings complying to graph topology while (2) the topology network trains to learn optimal embeddings with shared learning parameters for semi-supervised node classification. In addition, the content network enforces adversarial training on the topology network in order to balance node content and topology information for optimal node representation learning. The difference between our proposed CoGL with existing Graph convolutional Networks (GCN) \cite{kipf2016semi} is explained in Fig. \ref{fig:diff}.

In summary, our main contribution is twofold: 1) We propose modeling inconsistency and discrepancy between network node content and topology for optimal and robust graph embedding; 2) We propose \text{\nickname}, a novel GNN model that enables co-alignment leaning between content based network and the original topology network. 

\section{Related Work}
%This section briefly reviews related graph embedding methods and graph neural networks that have received significant attentions recently. 
Given a network with edge connections and content (features) associated to each node, graph embedding learns a low-dimensional vector for each node to preserve node content and network topology~\cite{shi2019topical}. Many works have been proposed, ranging from unsupervised learning methods such as DeepWalk \cite{perozzi2014deepwalk} to supervised learning methods such as SemiGraph \cite{hisano2018semi}. The common idea is that nodes with similar topology or similar content should be represented with similar embeddings in the latent space~\cite{cai2018comprehensive}.

\begin{figure*}
  \centering
    \includegraphics[width=0.9\textwidth]{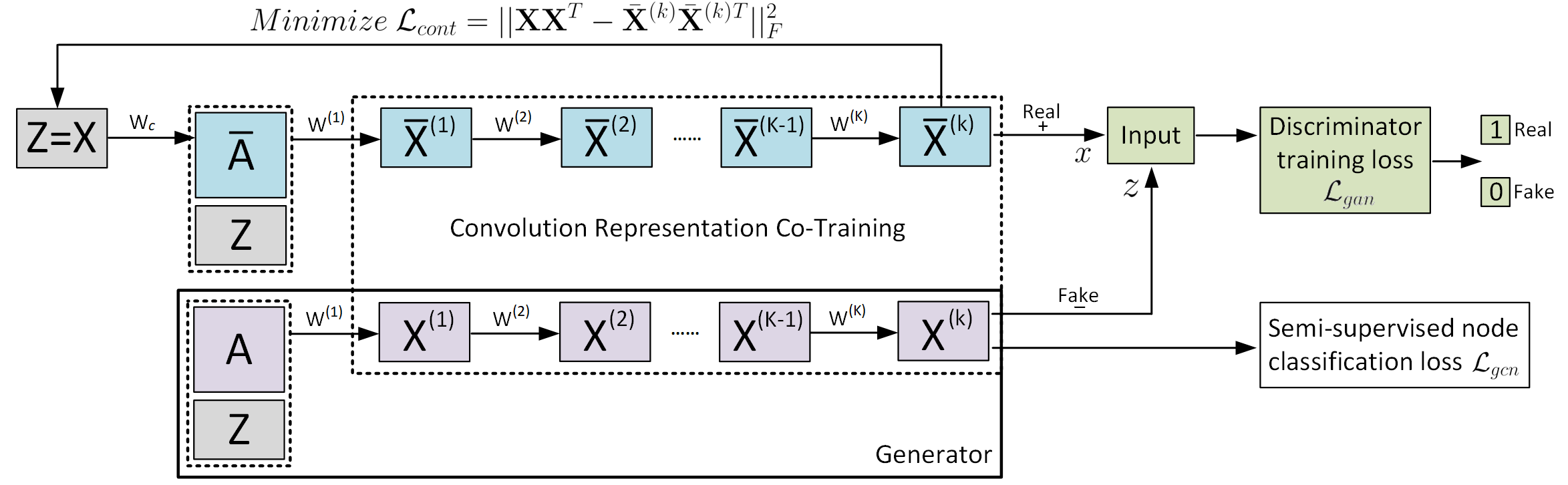}
  \caption{The proposed \text{\nickname} model. It performs two convolution representation learning on content network ($\adjmatbar$) and topology network ($\adjmat$), respectively. The content network tries to learn good embeddings that best reconstruct $\adjmatbar$ by minimizing $\lossfeat$, while the topology network tries to learn optimal embeddigs with co-trained parameters by optimizing node classification loss $\lossgcn$. Meanwhile, The content network enforces an adversarial training on the topology network by optimizing binary classification loss $\lossgan$.}
  \label{fig:cogl}
  \vspace{-2mm}
\end{figure*}

Graph Neural Networks (GNN) \cite{wu2019comprehensive} is a family of neural network models specifically developed for learning grid-like networked data. GNN models usually have efficient information aggregators \cite{du2019graph} that apply directly on graphs to easily incorporate graph structures and features for unified node representation learning. Graph Convolutional Networks (GCN) \cite{kipf2016semi} adopt a spectral-based convolution filter by which nodes can aggregate features from their respective local graph neighborhoods for representation learning. This convolution learning mechanism has been proven successful in many real-word analytic tasks such as link prediction \cite{zhang2018link}, image recognition \cite{chen2019multi} and new drug discovery \cite{sun2019graph}.  Following the similar convolution graph embedding scheme, many GNN models with more efficient information aggregators have been proposed. For example, Graph Attention Network (GAT) \cite{velivckovic2017graph} learns to assign different importance weights for various nodes so that nodes can highlight important neighborhoods while aggregating features. GraphSAGE \cite{hamilton2017inductive} learns a set of aggregation functions for each node to flexibly aggregate information from neighborhoods within different hops.

For all existing GNN approaches, graph convolutional learning is carried out by using topology to drive feature aggregation. A widely accepted assumption is that node content and graph structures are consistent and complementary for measuring node closeness in the embedding space~\cite{pan2016tri,maheshwari2019representation}. In reality, the network topology is often noisy and inconsistent to node content. A recent study~\cite{Guo2019cofond} shows that network topology is not only inconsistent to node content at the individual node level, but is also different from the affinity network (built from node content) at the network level (\textit{e.g.} degree distributions). A recent topology optimization based GCN~\cite{Yang2019topology} proposes to revise network topology to improve GCN learning, but requires revising/changing network topology which is risky to most users. 

%Following previous settings, we focus on the semi-supervised GNN model for information graph learning in this paper. 
Different from existing models that consider graph structures and node content as consistent by default or revise them to ensure consistency, we propose a new co-alignment paradigm to explicitly model their inconsistency for optimal and robust graph embedding.

\section{Problem Definition and Framework}
\subsection{Problem Definition}
Without loss of generality, an information graph can be represented as $\graph=(\nodes,\edges,\feats)$, where $\nodes=\{\nodei\}_{i=1,\cdots,\nodenum}$ is a set of unique nodes and $\edges=\{\edgeij \}_{ {i,j}=1,\cdots,\nodenum;~i \neq j }$ is a set of edges which can be equal to a $\nodenum \times \nodenum$ adjacency matrix $\adjmat$ with $\Aij=\wij>0$ if $\edgeij \in \edges$ and $\Aij=0$ if $\edgeij \notin \edges$. $\feats \in \R^{\nodenum \times \featdim}$ is a matrix containing all $\nodenum$ nodes with their features, $i.e.$, $\feati \in \R^\featdim$ represents the feature vector of node $v_i$, where $\featdim$ is the feature vector's dimension. It is easy to conclude that $\graph$ can be any type of networked data where nodes have feature contents, such as citation networks with word texts as node features and image networks with image semantics as node features. Given an information graph $\graph$, we aim to learn graph embeddings $\embeds \in \R^{\nodenum \times \embedsize}$ for node classification in a semi-supervised fashion, where $\embedsize$ is the embedding size. In this paper, the expressions of graph embedding and representation learning are used interchangeably.

\subsection{The Overall Framework}
The proposed \text{\nickname} model for semi-supervised graph embedding and node classification is shown in Fig. \ref{fig:cogl}. It performs a co-alignment learning between the constructed content network and the original topology network through three collaborative components:
\begin{itemize}
    \item \textbf{Content-aligned Graph Topology Learning:} This part trains to learn the content-aligned graph topology network $\adjmatbar$ (content network) from input node features $\feats$ by minimizing a graph reconstruction loss. 
    \item \textbf{Semi-supervised Graph Embedding:} This part trains to learn convolution graph embeddings from the original graph topology network (topology network) $\adjmat$ for node classification in a semi-supervised manner.  
    \item \textbf{Adversarial Graph Embedding Training:} In this part, the content network embedding enforces an adversarial training on the topology network embedding.
\end{itemize}

\section{The Proposed Model}
\subsection{Content-aligned Graph Topology Learning}
We aim to learn a non-negative matrix $\adjmatbar \in \R^{\nodenum \times \nodenum}$ (e.g., we also call the content network) that could reveal the underlying pairwise node relationships from node features $\feats$. An ad-hoc solution is to build a $k$-nearest neighbor graph \cite{maier2009influence} or simply calculate the Euclidean distance for each pair of nodes by $\Abarij=\left\lVert \feati-\featj \right\rVert$. However, this does not generate optimal graph embeddings for specific problems. Instead, we adopt a single-layer feed-forward neural network parametrized by a weight vector $\Wc \in \R^{\featdim \times 1}$ and followed by a non-linear transformation (e.g., $\relu(u)=\max(0,u)$). It takes the feature difference $|\feati-\featj|$ between $i^{th}$ and $j^{th}$ nodes as input and outputs the corresponding relevance weight by: 
\begin{equation}
    \wij = \relu(\Wc^T|\feati-\featj|)
\end{equation}
Then, we apply softmax normalization on each node and finally obtain the weight matrix $\adjmatbar$ as:

\begin{equation}
    \Abarij = \text{softmax}(\wij)= \frac{\text{exp}(\relu(\Wc^T|\feati-\featj|))}{\sum^{\nodenum}_{j=1}\text{exp}(\relu(\Wc^T|\feati-\featj|))}
\end{equation}
We can observe that $\adjmatbar$ is no longer a symmetric matrix (i.e., $\adjmatbar_{i,j}\neq\adjmatbar_{j,i}$) after the normalization, which is rational as the target is to learn an aligned approximation of nodes to their respective neighborhoods for optimal graph embedding. For efficient calculation, we can first map input node features ($\feats$) to a dimension-reduced space \cite{velivckovic2017graph,jiang2019semi} and Equation (2) can be written as:
\begin{equation}
    \Abarij = \frac{\text{exp}(\relu(\Wc^T|\feati\Wp-\featj\Wp|))}{\sum^{\nodenum}_{j=1}\text{exp}(\relu(\Wc^T|\feati\Wp-\featj\Wp|))}
\label{eq3}
\end{equation}
where $\Wp \in \R^{\featdim \times \wpdim}$ is a learned matrix and $\wpdim < \featdim$.

\begin{algorithm}[t]
\begin{small}
\SetAlgoLined
\SetKwInOut{Input}{Input}\SetKwInOut{Output}{Output}
\Input{Graph topology $\adjmat$ and node features $\feats$}
\Output{Node embeddings $\embeds \in \R^{\nodenum\times\embedsize}$}
\textbf{Initialization}: $i=0$, training epochs $M$ and $N$
\BlankLine
\While{$i \le M$}
{
    $\adjmatbar \leftarrow$ build content network through Eq.(\ref{eq3})\; 
    $\embedbar \leftarrow$ learn content network embeddings through Eq.(\ref{eq4})\;
    $\embeds \leftarrow$ learn topology network embeddings through Eq.(\ref{eq6})\;
    \For {$j=0,\cdots,N$}
    {
        Sample $n$ instances from content network embeddings $\embedbar$\;
        Sample $n$ instances from topology network embeddings $\embeds$\;
        Update model parameters based on Eq.(\ref{eq10}) in two steps: \\
        \begin{itemize}
            \item Optimize $\mathcal{L}$ while training the discriminator;
            \item Optimize $\mathcal{L}$ while training the generator.
        \end{itemize} 
    }
    $i = i + 1$.
}\caption{Training the \text{\nickname} model}
\end{small}
\end{algorithm}

To derive a consistent content network that best aligns node features, we first perform a two-layer convolutional representation learning \cite{kipf2016semi} based on $\adjmatbar$ by:
\begin{equation}
    \embedbar = \adjmatbar\relu(\adjmatbar\feats\Wone)\Wtwo
\label{eq4}
\end{equation}
where $\Wone \in \R^{\featdim \times \hiddendim}$ and $\Wtwo \in \R^{\hiddendim \times \embedsize}$ are the first and second convolution-layer embedding parameters, respectively. Then, we minimize the reconstruction error between learned node embeddings and input node features as:
\begin{equation}
    \lossfeat = \parallel \feats{\feats}^T-\embedbar{\tilde{\feats}}^{(2)T}\parallel^2_F
\label{eq5}
\end{equation}
where both $\feats$ and $\embedbar$ have been normalized to ensure stable parameter learning, \textit{i.e.}, through $\feats=\text{softmax}(\feats)$ and $\embedbar=\text{softmax}(\embedbar)$.

\subsection{Semi-supervised Graph Embedding}
We perform graph representation learning for node classification in a semi-supervised manner. In a similar way, the low-dimensional embeddings are derived through a two-layer recursive convolution learning based on the original topology network $\adjmat$ by:
\begin{equation}
    \embeds = \adjmattilde\relu(\adjmattilde\feats\Wone)\Wtwo
\label{eq6}
\end{equation}
where $\adjmattilde=\degmat^{-\frac{1}{2}}(\idenmat+\adjmat)\degmat^{-\frac{1}{2}}$ denotes the normalized form of $\adjmat$, $\idenmat$ is an identity matrix with the same shape and $\degmat$ is the degree matrix of $(\idenmat+\adjmat)$. Here, the convolution learning parameters $\Wone$ and $\Wtwo$ are shared and co-trained between the content network and the topology network, which are beneficial to balance node content and topology for graph construction and embedding learning. Then, we do a semi-supervised training to learn the parameters by minimizing the node classification loss $\lossgan$ as follows:
\begin{equation}
    \embedsoftmax = \text{softmax}(\embeds) = \frac{{\exp (\embeds )}}{{\sum\nolimits_t {\exp (\embeds_t)}}}
\label{eq7}
\end{equation}
\begin{equation}
    \lossgcn = -\sum\limits_{i\in \labnodes}\sum\limits_{j=1}^\embedsize\labonehot_{i,j}\ln\embedsoftmax_{i,j} 
\label{eq8}
\end{equation}
where the node embedding size $\embedsize$ equals the total number of labels for graph nodes. $\labonehot \in \R^{\nodenum \times \embedsize}$ denotes the one-hot label indicators matrix for all nodes and $\labnodes$ is a set of node indices with labels known for semi-supervised training.

\begin{table}
\centering
\caption{Document network characteristics.}
\begin{small}
\label{tab:data}
    \begin{tabular}{c|c|c|c|c}
    \hline
    Items & Cora & Citeseer & PubMed & DBLP \\
    \hline
    \# Nodes & $2708$  & $3327$ & $19717$ &$17725$ \\
    \hline
    \# Edges & $5429$ & $4732$ & $44338$ &$52890$ \\
    \hline
    \# Features & $1433$ & $3703$ & $500$ & $6974$ \\
    \hline
    \# Classes & $7$ & $6$ & $3$ & $4$\\
    \hline
    \end{tabular}
\end{small}
\vspace{-3mm}
\end{table}

\begin{table}
\centering
\caption{Image network characteristics.}
\begin{small}
\label{tab:data}
    \begin{tabular}{c|c|c|c|c}
    \hline
    Items (\#) & Nodes & Edges & Features & Classes \\
    \hline
    MIR & $5892$  & $380808$ & $500\times375$ &$152$ \\
    \hline
    ImageCLEF & $3461$ & $221185$ & $500\times375$ &$134$ \\
    \hline
    \end{tabular}
\end{small}
\vspace{-3mm}
\end{table}

\begin{table*}[ht]
\renewcommand{\arraystretch}{1.20}
\centering
\caption{Node classification accuracy on Cora, Citeseer, Pubmed and DBLP datasets.}
\label{tab:results}
\begin{tabular}{c|cccc}
\toprule
\textbf{Methods} & \textbf{Cora} & \textbf{Citeseer} & \textbf{Pubmed} & \textbf{DBLP} \\
\toprule
  DeepWalk & {$67.2\%$} & {$43.2\%$} & {$65.3\%$} &  {$66.3\%$} \\

 SemiEmd & {$59.0\%$} & {$59.6\%$} &   {$71.7\%$} & {$72.1\%$}  \\

 Planetoid & {$75.7\%$} & {$64.7\%$} &   {$77.2\%$} & {$74.7\%$}  \\

 Chebyshev & {$81.2\%$} & {$69.8\%$} &  {$74.4\%$} & {$74.3\%$} \\

 GCN & {$81.5\%$} & {$70.9\%$} &   {$79.0\%$} & {$75.1\%$} \\

 GAT & {$83.2\pm0.7\%$} & {$71.0\pm0.7\%$} & $79.0\pm0.3\%$ & {$78.2\pm0.7\%$}\\
\midrule
 \text{\nickname} (\textbf{Ours}) & {$\textbf{84.1}\pm0.6\%$} & {$\textbf{72.4}\pm0.5\%$} &   {$\textbf{79.2}\pm0.3\%$} & {$\textbf{79.8}\pm0.6\%$}  \\

\bottomrule
\end{tabular}
% \vspace{-5mm}
\end{table*}

\subsection{Adversarial Graph Embedding Training}
The learning balance of content and topology is achieved through adversarial training \cite{goodfellow2014generative}, where the goal is to consider both content and topology information for optimal graph embedding and node classification. As shown in Fig. \ref{fig:cogl}, the topology network learning component is considered a generator $\generator$ that generates node embeddings based on the topology network $\adjmat$. During adversarial training, the discriminator tries to classify the real node (positive) sample $\sample\in \embedbar$ learned from the content network $\adjmatbar$ as class 1 and meanwhile classify the fake node (negative) sample $\fakesample=\generator$ as class 0. On the contrary, the generator tries to fool the discriminator by classifying $\fakesample$ as class 1. The adversarial embedding training objective is given as:

\begin{equation}
\begin{aligned}
    \underset{\gen}{\min}~\underset{\dis}{\max}~\lossgan(\dis,\gen) = \E_{\realdatadist}\log\discriminator \\ 
    +\E_{\fakedatadist}\log(1-\dgnator) 
\end{aligned}
\label{eq9}
\end{equation}

\subsection{Model Optimization and Training}
As described above, the content network construction and graph embedding conform to an adversarial co-alignment learning fashion for optimal node classification performance. As the three components in Fig. \ref{fig:cogl} share and co-train convolution graph embedding parameters $\Wone$ and $\Wtwo$, we finally seek to optimize the following combined objective as:
\begin{equation}
    \mathcal{L} = \lossgcn + \alpha \lossfeat + \beta \lossgan(\dis,\gen)
\label{eq10}
\end{equation}
where $\alpha$ and $\beta$ are set to balance the content network construction and the adversarial embedding training, respectively. Note that $\lossgan(\dis,\gen)$ in Equation (9) involves both discriminator and generator training, where each part needs to combine both $\lossgcn$ and $\lossfeat$ for collective model parameter optimization. The training procedure of our \text{\nickname} model is summarized in Algorithm 1.

\section{Experiments}
In this section, we evaluate the proposed model for supervised node classification on six benchmark datasets. The dataset statistics are summarized in Tables 1 and 2.

% \vspace{2p}
\subsection{Datasets}
% \noindent \textbf{Datasets.} 
We use four document networks Cora, Citeseer, Pubmed and DBLP that have been widely used for node-level classification in previous work
\cite{kipf2016semi,leskovec2014snap}. \textbf{Cora} contains 2708 research papers grouped into 7 machine learning classes such as \textit{Reinforce Learning} and \textit{Genetic Algorithms}. There are 5429 edges between them and each paper node is described with a feature vector of 1433 dimensions. \textbf{Citeseer} contains 3327 research papers in 6 classes with 4732 links between them, where each paper node has a feature vector of 3703 dimensions. \textbf{Pubmed} contains 19717 literature nodes and 44338 edges. Each node belongs to one of the 3 classes and has a feature vector of 500 dimensions. \textbf{DBLP} contains 17725 publications from 4 classes. It has 52890 edges and each node is associated with a feature vector of 6974 dimensions.

We also use two multi-label image networks\footnote{https://snap.stanford.edu/data/web-flickr.html} MIR and ImageCLEF. \textbf{MIR} has 5892 nodes from 152 classes. Each node represents a $500\times375$ RGB color image and there are 380808 edges for this network. \textbf{ImageCLEF} contains 3461 nodes and 221185 edges. Each node is also a $500\times375$ RGB image that corresponds to one or more of the 134 classes. For each image in these two datasets, we extract a CNN feature descriptor and the feature dimensions for MIR and ImageCLEF are transformed to 152 and 134, respectively.

\begin{figure}
\begin{minipage}[b]{0.5\linewidth}
\centering 
\includegraphics[width=1\textwidth]{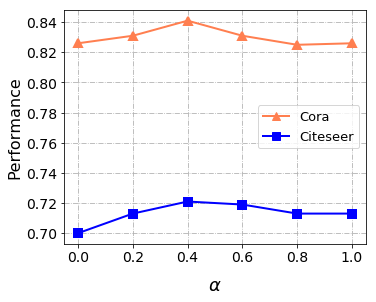}
\end{minipage}%
% \caption{Impact of the parameter $\alpha$.}
% \label{fig:3}
\begin{minipage}[b]{0.5\linewidth}
\centering 
\includegraphics[width=1\textwidth]{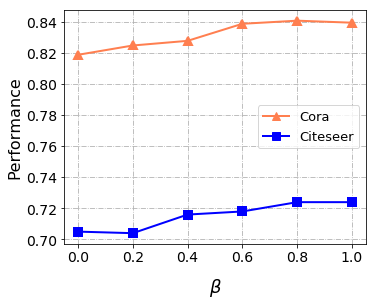}
\end{minipage}%
\caption{Impact of parameters $\alpha$ and $\beta$.}
\label{fig:4}
\vspace{-3mm}
\end{figure}

% \begin{figure}
% \begin{minipage}[b]{1\linewidth}
% \centering 
% \includegraphics[width=0.56\textwidth]{beta.png}
% \end{minipage}%
% \caption{Impact of the parameter $\beta$.}
% \label{fig:3}
% \end{figure}

%\vspace{2pt}
\subsection{Compared Methods and Settings}
\noindent \textbf{Compared Methods.} We compare our proposed \text{\nickname} model for semi-supervised node classification against various state of the arts, including the random walk-based unsupervised embedding method (\textbf{DeepWalk}) \cite{perozzi2014deepwalk}, the semi-supervised embedding methods \textbf{SemiEmb} \cite{weston2012deep} and \textbf{Planetoid} \cite{yang2016revisiting}, graph neural network models such as the spectrum-based graph convolutional networks (\textbf{GCN}) \cite{kipf2016semi}, the \textbf{Chebyshev} \cite{defferrard2016convolutional} filter-based GCN and the graph attention networks (\textbf{GAT}) \cite{velivckovic2017graph} that additionally take importance of neighborhoods into consideration.

\begin{table*}[ht]
\renewcommand{\arraystretch}{1.20}
\centering
\caption{Node classification accuracy (\%) on MIR and ImageCLEF datasets.}
\label{tab:results}
\begin{tabular}{c|ccc|ccc}
\toprule
\multicolumn{1}{c|} {\textbf{Datasets}} & 
\multicolumn{3}{c}{\textbf{MIR}} & 
\multicolumn{3}{|c}{\textbf{ImageCLEF}} \\
\midrule
\multicolumn{1}{c|}{\# \textbf{Labeled Nodes}} & 300 & 500 & 700 & 300 & 500 & 700\\ \toprule
  DeepWalk & {$41.42\pm0.22$} & {$42.98\pm0.25$} & {$44.50\pm0.33$} &  {$43.34\pm0.50$} &  {$43.67\pm0.39$}&  {$44.98\pm0.43$}\\

 SemiEmd & {$46.75\pm0.39$} & {$47.24\pm0.48$} & {$48.61\pm0.27$} &  {$54.30\pm0.67$} &  {$55.22\pm0.49$}&  {$56.50\pm0.42$}  \\

 Planetoid & {$50.74\pm0.33$} & {$51.04\pm0.40$} & {$51.91\pm0.23$} &  {$56.75\pm0.36$} &  {$57.77\pm0.36$}&  {$57.99\pm0.49$}  \\

 Chebyshev & {$53.28\pm0.32$} & {$54.28\pm0.32$} & {$55.87\pm0.45$} &  {$58.12\pm0.37$} &  {$58.79\pm0.51$}&  {$59.58\pm0.51$} \\

 GCN & {$55.43\pm0.41$} & {$56.14\pm0.38$} & {$57.65\pm0.27$} &  {$60.33\pm0.34$} &  {$61.04\pm0.54$}&  {$61.60\pm0.38$} \\

 GAT & {$55.59\pm0.54$} & {$57.27\pm0.76$} & {$58.38\pm0.48$} &  {$61.45\pm0.52$} &  {$61.47\pm0.32$}&  {$62.32\pm0.41$}\\
\midrule
 \text{\nickname} (\textbf{Ours}) & {$\textbf{57.23}\pm\textbf{0.46}$} & {$\textbf{58.78}\pm\textbf{0.39}$} & {$\textbf{60.11}\pm\textbf{0.45}$} &  {$\textbf{61.77}\pm\textbf{0.57}$} &  {$\textbf{62.52}\pm\textbf{0.40}$}&  {$\textbf{63.86}\pm\textbf{0.52}$}  \\

\bottomrule
\end{tabular}
% \vspace{-5mm}
\end{table*}

\begin{figure*}[!ht]
   \begin{minipage}{0.50\textwidth}
     \centering
     \includegraphics[width=0.45\linewidth]{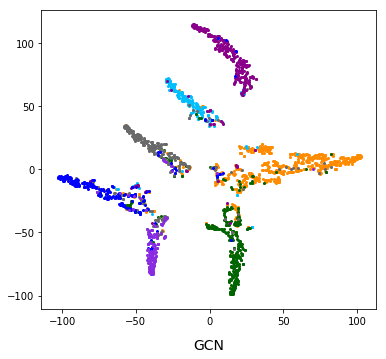}% 
     \includegraphics[width=0.45\linewidth]{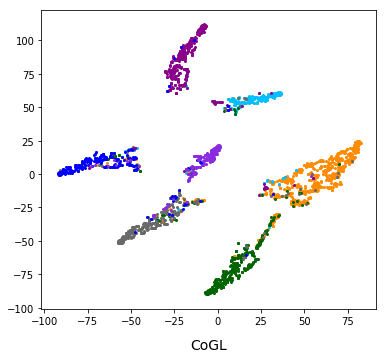}%
     \caption{ Embedding 2-D visualization on Cora dataset.}
   \end{minipage}
   \hfill
   \begin{minipage}{0.5\textwidth}
     \centering
     \includegraphics[width=0.525\linewidth]{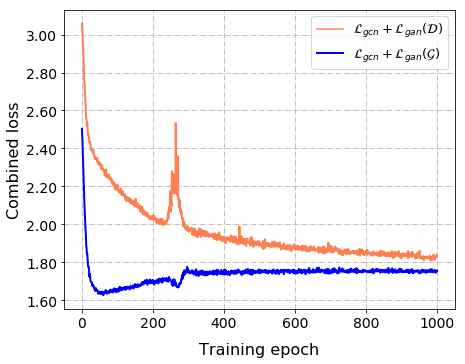}
     \caption{Classification training loss.}
   \end{minipage}
\vspace{-3mm}
\end{figure*}

%\vspace{2pt}
\noindent \textbf{Experimental Settings.} For experiments of baseline methods on Cora, Citeseer, Pubmed and DBLP datasets, we follow the same settings as in previous works \cite{kipf2016semi,velivckovic2017graph}. For the MIR and ImageCLEF datasets, we respectively select 300, 500 and 700 labeled image nodes for training the model. For other unlabeled nodes, we select 1000 and 2000 images for validation and test, respectively. For each experiment, data splits are performed randomly and the average performances are finally reported over ten runs. 

In our approach, the hidden layer dimensions $d$ for content network construction is set as 70 and the hidden convolution layer dimension $h$ is set as 30. We train \text{\nickname} for a maximum of 1000 epochs based on the Adam algorithm with an early stopping of 200 epochs. For the four document networks, the dropout probability and learning rate are set as 0.5 and 0.002, and they are 0.2 and 0.01 for MIR and ImageCLEF datasets. For comparison, we set the balance parameters $\alpha$ and $\beta$ as 0.4 and 0.8, respectively. Similar to previous work \cite{kipf2016semi}, we use a $L_2$ norm regularization where the weight decay is set as 5e-4.

% \vspace{2pt}
\subsection{Node Classification Results}
Table 3 shows the node classification results on Cora, Citeseer, Pubmed and DBLP datasets, and the results for MIR and ImageCLEF datasets are shown in Table 4. The best results in the two tables are highlighted. We can observe that our \text{\nickname} model outperforms other GNN models such as GCN and GAT for semi-supervised node classification. Existing GNN models naturally consider network structures and node features as two unbiased sources of data where each part consistently reveals and enhances the another during learning. However, the topology-based node relations are not always matched with the content-based node relations due to the existence of irrelevant and missing links between nodes in the original graph. Our method can model this discrepancy with the co-alignment learning between the content and topology networks. The experimental results of \text{\nickname} compared with existing GNN models and other state-of-the-art baselines such as SemiEmd and DeepWalk demonstrates the effectiveness of our approach for information network learning. In addition, the visualization results in Fig. 5 show that \text{\nickname} can learn more discriminative node embeddings compared with GCN.

% In addition, for multi-label node classification results on the two image networks in Table 4, we can observe the proposed \text{\nickname} model still performs better than other baselines with different number of labeled nodes for supervised training. This again verifies that it is beneficial to perform adversarial co-alignment learning between content network and topology network for some graph data.

% \begin{figure}[t]
% \centering
% \begin{minipage}[b]{0.6\linewidth}
% \centering 
% \includegraphics[width=1\textwidth]{gcn_cora.png}
% \end{minipage}%
% \caption{The label correlation matrix.}
% \label{fig:6}
% \vspace{-5mm}
% \end{figure}

% \begin{figure}[t]
% \centering
% \begin{minipage}[b]{0.6\linewidth}
% \centering 
% \includegraphics[width=1\textwidth]{our_cora.png}
% \end{minipage}%
% \caption{The label correlation matrix.}
% \label{fig:6}
% \vspace{-5mm}
% \end{figure}

\vspace{-2mm}
\subsection{Parameter Sensitivity and Classification Loss}
The impacts of parameters $\alpha$ and $\beta$ are shown in Fig. \ref{fig:4}. We observe that either ignoring the content network construction component ($\alpha=0.0$) or ignoring the adversarial training component ($\beta=0.0$) during training causes the lowest classification performance, which verifies the benefit of combining content network and topology network for co-alignment training in this paper. 

To demonstrate the balanced learning between topology network and content network, we show the combined loss of node classification ($\lossgcn$) and adversarial binary classification (discriminator $\lossgan(\mathcal{D})$ or generator $\lossgan(\mathcal{G})$) in Fig. 6. We observe that the two curves tend to be closer but there still a difference, which reflects an adversarial balance between topology network and content network for unified and optimal graph embedding and node classification. 

%\vspace{2pt}

\vspace{-3mm}
\section{Conclusion} 
%Graph content and structure are two main sources of data that have been widely used in many graph embedding models. However, current methods naturally consider structure and content exist to reflect unbiased relationships between nodes, which is not always the case in reality. In this paper, we proposed taking the inconsistency between graph structure and features into consideration for balanced and robust graph embedding learning. A novel GNN model, \text{\nickname}, is introduced accordingly. The merit of \text{\nickname} lies in that the original topology network can co-align the constructed content network for unified and optimal node embedding and classification. Experiments and validations on six benchmark datasets demonstrated the effectiveness of \text{\nickname}. The proposed model can be used to learn embeddings for all types of information networks where nodes have features such as the widely seen document networks and image networks. %We tested the sensitivity of various parameters involved and demonstrated how correlated labels interact with each other thereby affecting classification performance.
Network topology and node content (including labels) are two main sources of information commonly used in many network embedding models. Although topology and node content are often inconsistent, existing methods, especially graph convolution networks, naturally ignore such inconsistency for embedding learning. In this paper, we took the inconsistency between graph structure and features into consideration for robust graph embedding learning, and proposed a novel co-alignment graph convolutional learning model, \text{\nickname}. The merit of \text{\nickname} lies in that it co-aligns the original topology network and the constructed content network for optimal node embedding and classification. Experiments and validations on six benchmark datasets demonstrated the effectiveness of \text{\nickname}. The proposed model can be used to learn embeddings for a variety of information networks with rich node content, such as the widely seen document networks and image networks. %We tested the sensitivity of various parameters involved and demonstrated how correlated labels interact with each other thereby affecting classification performance.

%\section*{Acknowledgments}

%\appendix

%\section{Appendix}\label{appendix-a}

%% The file named.bst is a bibliography style file for BibTeX 0.99c
\bibliographystyle{IEEEtran}
\bibliography{ijcai20}

% Generated by IEEEtran.bst, version: 1.14 (2015/08/26)
\begin{thebibliography}{10}
\providecommand{\url}[1]{#1}
\csname url@samestyle\endcsname
\providecommand{\newblock}{\relax}
\providecommand{\bibinfo}[2]{#2}
\providecommand{\BIBentrySTDinterwordspacing}{\spaceskip=0pt\relax}
\providecommand{\BIBentryALTinterwordstretchfactor}{4}
\providecommand{\BIBentryALTinterwordspacing}{\spaceskip=\fontdimen2\font plus
\BIBentryALTinterwordstretchfactor\fontdimen3\font minus
  \fontdimen4\font\relax}
\providecommand{\BIBforeignlanguage}[2]{{%
\expandafter\ifx\csname l@#1\endcsname\relax
\typeout{** WARNING: IEEEtran.bst: No hyphenation pattern has been}%
\typeout{** loaded for the language `#1'. Using the pattern for}%
\typeout{** the default language instead.}%
\else
\language=\csname l@#1\endcsname
\fi
#2}}
\providecommand{\BIBdecl}{\relax}
\BIBdecl

\bibitem{fan2019graph}
W.~Fan, Y.~Ma, Q.~Li, Y.~He, E.~Zhao, J.~Tang, and D.~Yin, ``Graph neural
  networks for social recommendation,'' in \emph{The World Wide Web
  Conference}.\hskip 1em plus 0.5em minus 0.4em\relax ACM, 2019, pp. 417--426.

\bibitem{chen2019multi}
Z.-M. Chen, X.-S. Wei, P.~Wang, and Y.~Guo, ``Multi-label image recognition
  with graph convolutional networks,'' in \emph{Proc. of IEEE CVPR}, 2019, pp.
  5177--5186.

\bibitem{kipf2016semi}
T.~N. Kipf and M.~Welling, ``Semi-supervised classification with graph
  convolutional networks,'' \emph{arXiv preprint arXiv:1609.02907}, 2016.

\bibitem{hamilton2017inductive}
W.~Hamilton, Z.~Ying, and J.~Leskovec, ``Inductive representation learning on
  large graphs,'' in \emph{Advances in Neural Information Processing Systems},
  2017, pp. 1024--1034.

\bibitem{velivckovic2017graph}
P.~Veli{\v{c}}kovi{\'c}, G.~Cucurull, A.~Casanova, A.~Romero, P.~Lio, and
  Y.~Bengio, ``Graph attention networks,'' \emph{arXiv preprint
  arXiv:1710.10903}, 2017.

\bibitem{wu2019comprehensive}
Z.~Wu, S.~Pan, F.~Chen, G.~Long, C.~Zhang, and P.~S. Yu, ``A comprehensive
  survey on graph neural networks,'' \emph{arXiv preprint arXiv:1901.00596},
  2019.

\bibitem{hofmann2017dbkwik}
A.~Hofmann, S.~Perchani, J.~Portisch, S.~Hertling, and H.~Paulheim, ``Dbkwik:
  Towards knowledge graph creation from thousands of wikis.'' in
  \emph{International Semantic Web Conference (Posters, Demos \& Industry
  Tracks)}, 2017.

\bibitem{liu2019projfe}
H.~Liu, L.~Bai, X.~Ma, W.~Yu, and C.~Xu, ``Projfe: Prediction of fuzzy entity
  and relation for knowledge graph completion,'' \emph{Applied Soft Computing},
  vol.~81, p. 105525, 2019.

\bibitem{bojcheski2018adversarial}
A.~Bojcheski and S.~G{\"u}nnemann, ``Adversarial attacks on node embeddings,''
  \emph{arXiv preprint arXiv:1809.01093}, 2018.

\bibitem{Yang2019topology}
L.~Yang, Z.~Kang, X.~Cao, D.~Jin, B.~Yang, and Y.~Guo, ``Topology optimization
  based graph convolutional network,'' in \emph{Proc. of the 28th IJCAI}, 2019.

\bibitem{shi2019topical}
M.~Shi, Y.~Tang, X.~Zhu, J.~Liu, and H.~He, ``Topical network embedding,''
  \emph{Data Mining and Knowledge Disc.}, pp. 1--26, 2019.

\bibitem{perozzi2014deepwalk}
B.~Perozzi, R.~Al-Rfou, and S.~Skiena, ``Deepwalk: Online learning of social
  representations,'' in \emph{Proc. of ACM SIGKDD}.\hskip 1em plus 0.5em minus
  0.4em\relax ACM, 2014, pp. 701--710.

\bibitem{hisano2018semi}
R.~Hisano, ``Semi-supervised graph embedding approach to dynamic link
  prediction,'' in \emph{International Workshop on Complex Networks}.\hskip 1em
  plus 0.5em minus 0.4em\relax Springer, 2018, pp. 109--121.

\bibitem{cai2018comprehensive}
H.~Cai, V.~W. Zheng, and K.~C.-C. Chang, ``A comprehensive survey of graph
  embedding: Problems, techniques, and applications,'' \emph{IEEE Trans. on
  KDE}, vol.~30, no.~9, pp. 1616--1637, 2018.

\bibitem{du2019graph}
S.~S. Du, K.~Hou, R.~R. Salakhutdinov, B.~Poczos, R.~Wang, and K.~Xu, ``Graph
  neural tangent kernel: Fusing graph neural networks with graph kernels,'' in
  \emph{Advances in Neural Information Processing Systems}, 2019, pp.
  5724--5734.

\bibitem{zhang2018link}
M.~Zhang and Y.~Chen, ``Link prediction based on graph neural networks,'' in
  \emph{Advances in Neural Information Processing Systems}, 2018, pp.
  5165--5175.

\bibitem{sun2019graph}
M.~Sun, S.~Zhao, C.~Gilvary, O.~Elemento, J.~Zhou, and F.~Wang, ``Graph
  convolutional networks for computational drug development and discovery,''
  \emph{Brief. in bioinformatics}, 2019.

\bibitem{pan2016tri}
S.~Pan, J.~Wu, X.~Zhu, C.~Zhang, and Y.~Wang, ``Tri-party deep network
  representation,'' \emph{Proc. of IJCAI}, 2016.

\bibitem{maheshwari2019representation}
A.~Maheshwari, A.~Goyal, A.~Kumar, M.~K. Hanawal, and G.~Ramakrishnan,
  ``Representation learning on graphs by integrating content and structure
  information,'' in \emph{2019 11th International Conference on Communication
  Systems \& Networks (COMSNETS)}.\hskip 1em plus 0.5em minus 0.4em\relax IEEE,
  2019, pp. 88--94.

\bibitem{Guo2019cofond}
T.~Guo, S.~Pan, X.~Zhu, and C.~Zhang, ``Cfond: Consensus factorization for
  co-clustering networked data,'' \emph{IEEE trans. on KDE}, 2019.

\bibitem{maier2009influence}
M.~Maier, U.~V. Luxburg, and M.~Hein, ``Influence of graph construction on
  graph-based clustering measures,'' in \emph{Advances in neural information
  processing systems}, 2009, pp. 1025--1032.

\bibitem{jiang2019semi}
B.~Jiang, Z.~Zhang, D.~Lin, J.~Tang, and B.~Luo, ``Semi-supervised learning
  with graph learning-convolutional networks,'' in \emph{Proc. of IEEE CVPR},
  2019, pp. 11\,313--11\,320.

\bibitem{goodfellow2014generative}
I.~Goodfellow, J.~Pouget-Abadie, M.~Mirza, B.~Xu, D.~Warde-Farley, S.~Ozair,
  A.~Courville, and Y.~Bengio, ``Generative adversarial nets,'' in
  \emph{Advances in neural information processing systems}, 2014, pp.
  2672--2680.

\bibitem{leskovec2014snap}
J.~Leskovec and A.~Krevl, ``Snap datasets: Stanford large network dataset
  collection,'' 2014.

\bibitem{weston2012deep}
J.~Weston, F.~Ratle, H.~Mobahi, and R.~Collobert, ``Deep learning via
  semi-supervised embedding,'' in \emph{Neural Networks: Tricks of the
  Trade}.\hskip 1em plus 0.5em minus 0.4em\relax Springer, 2012, pp. 639--655.

\bibitem{yang2016revisiting}
Z.~Yang, W.~W. Cohen, and R.~Salakhutdinov, ``Revisiting semi-supervised
  learning with graph embeddings,'' \emph{arXiv preprint arXiv:1603.08861},
  2016.

\bibitem{defferrard2016convolutional}
M.~Defferrard, X.~Bresson, and P.~Vandergheynst, ``Convolutional neural
  networks on graphs with fast localized spectral filtering,'' in
  \emph{Advances in neural information processing systems}, 2016, pp.
  3844--3852.

\end{thebibliography}

\vfill

\end{document}